# Robust cryogenic matched low-pass coaxial filters for quantum computing applications

**Andrey A. Samoylov[1,2], Anton I. Ivanov[1,2], Vladimir V. Echeistov[1,2], Elizaveta I. Malevannaya[1,2], Aleksei R. Matanin[1,2], Nikita S. Smirnov[1,2], Victor I. Polozov[1,2], and Ilya A. Rodionov[1,2,a]**.

[1]FMN Laboratory, Bauman Moscow State Technical University, Moscow 105005, Russia
[2]Dukhov Automatics Research Institute (VNIIA), Moscow 127055, Russia

[a]Author to whom correspondence should be addressed: irodionov@bmstu.ru

Electromagnetic noise is one of the key external factors decreasing superconducting qubits coherence. Matched coaxial filters can prevent microwave and IR photons negative influence on superconducting quantum circuits. Here, we report on design and fabrication route of matched low-pass coaxial filters for noise-sensitive measurements at milliKelvin temperatures. A robust transmission coefficient with designed linear absorption (-1dB/GHz) and ultralow reflection losses less than -20 dB up to 20 GHz is achieved. We present a mathematical model for evaluating and predicting filters transmission parameters depending on their dimensions. It is experimentally approved on two filters prototypes different lengths with compound of Cu powder and Stycast commercial resin demonstrating excellent matching. The presented design and assembly route are universal for various compounds and provide high repeatability of geometrical and microwave characteristics. Finally, we demonstrate three filters with almost equal reflection and transmission characteristics in the range from 0 to 20 GHz, which is quite useful to control multiple channel superconducting quantum circuits.

Modern superconducting quantum processors and simulators suffer from extremely high sensitivity of their logical elements, qubits, to external noise. There are plenty of factors affecting the qubits performance, but one of the dominants are noisy electromagnetic and IR fields [2],[3]. It can be originated from both external sources, like thermal radiation or electromagnetic field generated by experimental setup elements, and internal sources, such as flux and charge noise. For external noise eliminating a special multilevel shielding is typically used [4]. To protect superconducting qubits from internal noise signal lines filters are required. The choice of the filter is usually determined by its amplitude frequency response and the signal line assignment for qubits control (Fig. 1). The qubit operation frequency is usually in the range from 4 GHz to 8 GHz, which imposes a condition on the filters that can be used for the qubit excitation line. Flux qubit lines are used for applying DC-bias and, they should ideally suppress the whole frequency range except DC (Fig. 1). Thus, microwave filters play integral role in the measurement circuit of quantum chips.

There is a number of microwave electronic components including wide range of LPF, HPF and BPF commercially available on the market [6]. These filters can be effectively used for excitation lines in different qubit measurement circuits [6], [7]. However, none of them may completely be used for flux control lines of the qubit, because of their undamped amplitude-frequency response in the passband. For this reason, experimental developments of handmade filters for various ranges are used. Modern handmade filters have reflection parameters of around -10dB, and various values of the transmission parameter [1]. There are a great variety of different filters implementations for noise suppression [6], [8], and one of them is a powder filter [9], [10], [11], [12]. It has a simple construction, available compound materials, which provides stable characteristics at milliKelvin temperatures [13], and easy to fabricate.

In this work, we report on design, assembly technology and cryogenic microwave characterization of the robust cryogenic impedance-matched low-pass coaxial powder filters. The proposed construction of the filters is universal and can be used in combination with any commercially available compounds. Robust transmission coefficient with designed linear absorption (-1dB/GHz) and low reflection losses (less than -20 dB up to 20 GHz) have been achieved using commercial compound ECCOSORB CR-110.

The base construction of the coaxial powder filter (Fig. 3a) has a complicated internal structure – suspended particles around the coil with an alternating signal flowing through it affect the emergent induction in a complex manner. In this case the filter impedance calculation to ensure matching with external circuit becomes a challenging task [14]. However, there are various powder filter experimental implementations and add-ons of the base construction: the winding type [13], electric capacitance addition for parametric increasing of damping [10], central wire straightening [14]. Basically, all the implementations with the coil as a central wire are very difficult for filter parameters direct calculation.

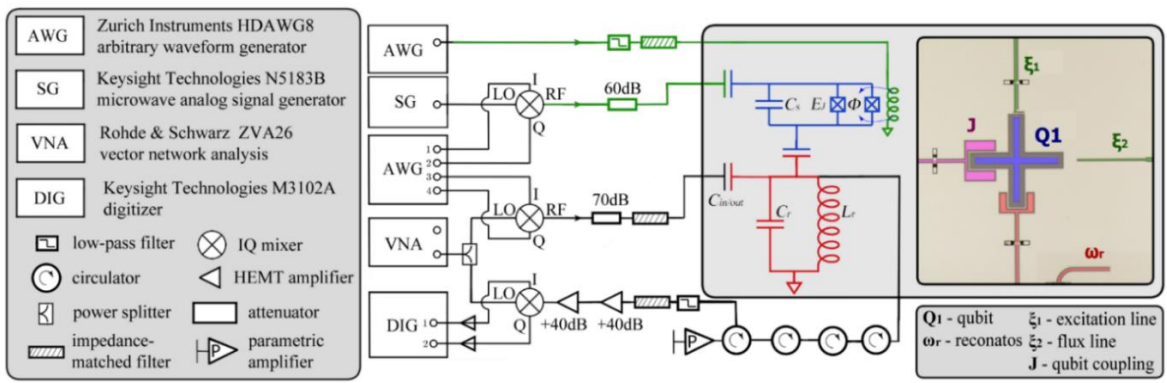

Fig.1. Multichannel measurement circuit of qubit-transmons. Optical photograph of X-mon: Qubit (blue) capacitively coupled by capacitive coupling element (violet) with another qubit. Qubit state is controlled by control lines (green): excitation line and flux line, readout via resonator $\omega_r$ (resonator)

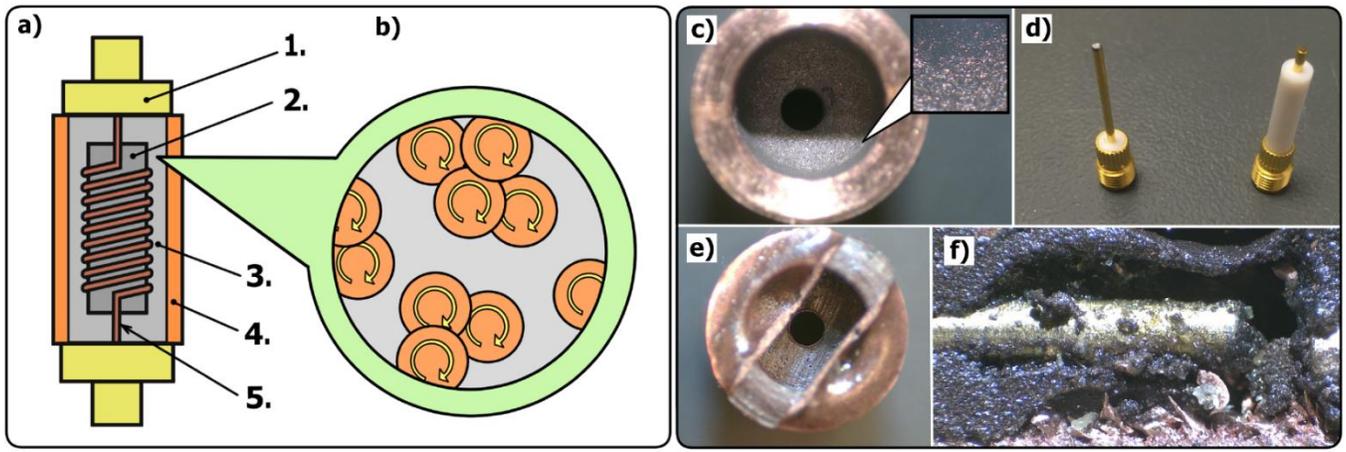

Fig.2. The base construction structure of the powder filter: a) sketching of the filter construction: 1 – SMA-connector; 2 – a core made of compound; 3 – outer filled compound; 4 – central wire as a coil; 5 – copper body. b) sketching of compound structure – metal particles suspended in epoxy resin; inside the particles inductive currents while the signal goes through the filter are originated; c) Filter blank with solidified compound of copper powder and epoxy resin; d) modified SMA-connector 142-1721-051 Johnson Components (left) and SMA-connector (right); e) Open curing of compound at epoxy resin temperature conditions; f) Compound curing at epoxy resin temperature conditions inside closed filter body

The fundamental concepts of powder filters [10] state that signal attenuation arises from the coil (as a central wire) and metal particles compound dissipates high-frequency signals. A coil as a central wire in the filter cylindrical body (Fig. 2a) creates additional induction which arises the attenuation when the signal goes through the filter [10]. Compound structure (Fig. 2b) – a mixture of epoxy resin and metal powder enables signal attenuation by inductive currents originating in the powder particles [10]. Thus, one can develop a powder filter with the simpler structure with its impedance dependent only on the properties of the effective "dielectric". Precise filter impedance control helps to minimize signal reflection from the ports, which can become new sources of noise. These powder filters with precise impedance control and minimized reflections are called matched powder filters [12].

Generally, the construction of the matched powder filter is the same as coaxial transmission line of finite length. The size of metal particles (~50 μm) in effective "dielectric" is small compared to the microwave wavelength in the working range, so the compound can be considered uniform (Fig. 2a,b). Mathematical equations for S-parameters of such finite length coaxial structures have been described before [15]:

$$\begin{cases} S_{21} = \dfrac{2}{2ch(\gamma l) + sh(\gamma l) * \left(\dfrac{Z}{Z_0} + \dfrac{Z_0}{Z}\right)} \\ S_{11} = \dfrac{\left(\dfrac{Z}{Z_0} - \dfrac{Z_0}{Z}\right)}{2cth(\gamma l) + \left(\dfrac{Z}{Z_0} + \dfrac{Z_0}{Z}\right)} \end{cases} \quad (1)$$

$$Z = \frac{1}{2\pi}\sqrt{\frac{\mu\mu_0}{\varepsilon\varepsilon_0}} \ln\frac{D}{d} \quad (2)$$

$Z$ – characteristic filter impedance, $Z_0$ – characteristic impedance of an empty waveguide, $\gamma = \alpha + i\omega\frac{\sqrt{\varepsilon\mu}}{c}$ - propagation constant, $\alpha$ – loss constant, l – transmission line length, d – inner diameter of coaxial line, D – outer diameter of coaxial line, μ - relative magnetic permeability, ε - relative dielectric constant, $\mu_0$ – permeability of free space, $\varepsilon_0$ – permittivity of free space.

Given equations completely describe the filter operation principle in the frequency domain, they can be used with the presence of effective filter parameters data ($\sqrt{\varepsilon\mu}$). These parameters are usually obtained from experimental data of a filter with arbitrary dimensions (outer and inner diameters, filter length) and the same compound. To confirm the validity of (1)-(2), in the framework of our research two filters of different lengths have been fabricated. Then we measured data for the first filter and used it for precise parameters calculation of the second one.

We started from a copper oxygen-free solid blank with the ratio of outer to inner diameters of 8:5,1 for filter body. We selected the round coaxial symmetrical body cross-section of the blank to guarantee robust characteristics of the filter [13], and oxygen-free copper as it serves better at cryogenic temperatures [16]. Inner diameter of the filter corresponds to the outer diameter of 142-1721-051 Johnson Components connector for further press fitting, this connector also easy-to-use for performing modifications (Fig. 2d). The body length is limited by the length of SMA-connector used. Fabricated filters had length of 42 mm and 36 mm. Inner diameter of coaxial filter is also limited by connector central wire. One can noticed that 142-1721-051 Johnson Components SMA-connector has a complicated construction. As central wire mechanical treatment leads to unrecoverable damage, the diameter of central wire does not change.

We choose STYCAST 2850FT as epoxy resin for compound with LOCTITE CAT 23LV as curing compound, which ensure appropriate thermal properties [14]. Nanografi copper powder with characteristic size of 50 μm was used as a metal powder [17]. We used the copper powder due to its magnetic properties – compound with copper powder will not create parasitic magnetic fields allowing to place the filters almost in any part of experimental setup even close to sensitive quantum circuits.

At the next fabrication step the copper body is filled in with the compound with copper powder and thermally treated. We found out that in copper body the compound is not solidified at the temperature specified in the datasheet, if it have no direct access to the ambient (Fig. 2f). There are two ways of a compound pouring into the body [18]: via side holes after press fitting or pouring with press fitting in uncured filler. Pouring via side holes of the filter body is more useful for fabrication, however, in this case the geometrical symmetry of the filter is missed, resulting in originating parasitic resonances [13]. To avoid this issue, we proposed to cure compound before press fitting the filters. A special teflon tooling was developed for it (Fig. 3a). Teflon was chosen due to its adhesion properties relative the epoxy resin used. During the thermal processing epoxy resin does not adhere to the tooling surface.

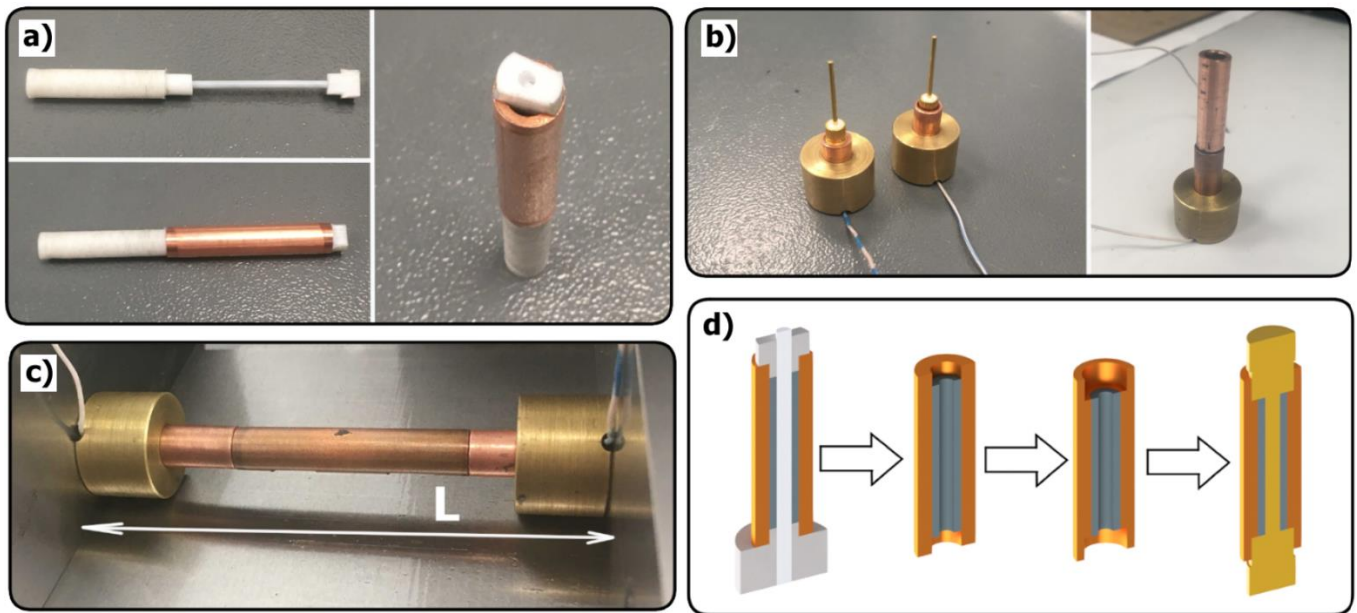

Fig. 3. a) Tooling for compound curing: assembled tooling for curing without filter, assembled tooling for curing covering filter body and top view; b) Tooling for press fitting: placing for support to connectors and placing for support to the body with pressed connector; c) Connectors press fitting to the filter blank. Linear dimensions controlled by distance measurement between vice plates L and its changes d) Filter assembly: all stages sequentially: compound curing, filter blank processing and connectors press fitting;

After thermal processing, all the teflon tooling are extracted and the shape of hardened compound is corrected by drilling. The grooves are drilled in filter blank in such a manner, that SMA-connectors during press fitting will be able to touch each other inside the filter. To guarantee the contact between connectors inside the filters during the further thermal processing (Table 1), the ends of the modified SMA-connectors are tinned by homemade low-melting solder with 63ºC melting temperature. Finally, the press fitting is performed.

To align the central rod of the connectors inside the filter body, the press fitting is performed by means of homemade assembly tooling (Fig. 3b). This tooling is designed to be able to press fitting SMA-connectors independently of each other and also to control connector contact during pressing. When the connectors touch each other, they complete an electrical circuit, which is recorded by multimeter.

| 16-24 hours | 25ºC |
| --- | --- |
| 4-6 hours | 45ºC |
| 2-4 hours | 65ºC |

Table. 1. Thermal processing of STYCAST 2850FT epoxy resin with LOCTITE CAT 23LV curing

Symmetrical output microwave characteristics of the filter are provided by linear size control from both sides of the filter (Fig. 3c). Moreover, it is necessary to achieve firm adherence of connectors tinned ends without damaging their central wire. To ensure it the hardened solder droplets on connectors are sharpened (Fig. 2d).

Transmission coefficient (S21) dependence on filter length (1)-(2), has been experimentally measured during the first stage of the filter fabrication development (Fig. 4a). Two filters of different lengths were made, filled with a mixture of epoxy and copper powder. Obtained parameters allowed us to calculate the transmittance S21 of a shorter filter with the tolerance of 5-10%. Reflection coefficient S11 has been calculated with the bigger error, since the model does not take into account the presence of side holes for pouring. It is also likely, that the difficulty of S11 calculation is related to compound imperfection with respect to considered model: a big amount of powder additionally shields the signal, which is ignored in the model.

Despite described above problems, filters construction has been modified. To remove the effect of filler imperfections, the fabrication process has been tested for the filters based on another filler – ECCOSORB CR-110 (Fig. 4b). In contrast to previous samples, the reflection is small enough (less than -20 dB), and attenuation values are repeatable with the high accuracy (in tolerance less than 5%). Nowadays, presented reflection parameters of our filters surpass known analogues [1] and is competitive with modern commercial filters [6]. The filters themselves are successfully used in cryogenic measuring systems [19], [20], [21], [22].

In conclusion, we studied the fabrication of matched co-axial filters for milliKelvin temperatures. In the course of the work, a mathematical model was proposed (1)-(2) that describes the dependence of S-parameters on the characteristics of the filler and the geometric dimensions of the filter. This model was confirmed (Fig 4a), and the detected inaccuracies in the determination of the S11 parameter were taken into account in the further development of the filters. For the final filter manufacturing technology (Fig 3d), special tooling has been developed for curing the filler (Fig 3a), as well as special tooling for pressing (Fig 3b) of modified SMA connectors (Fig 2d).

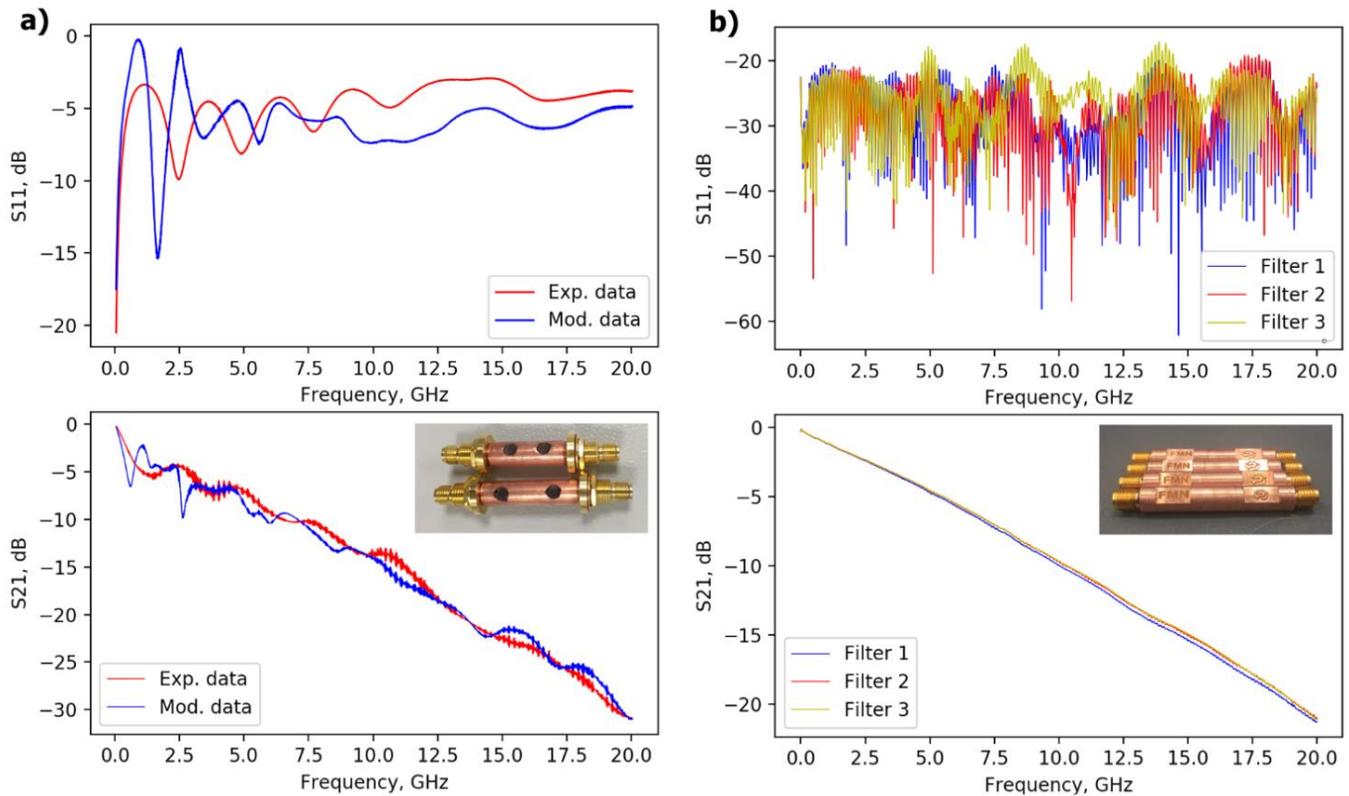

Fig. 4. a) Powder filters: 36-mm filter S-parameters: experimental data (Exp. data); and calculated data (Calc. data) based on the data of 42-mm filter; b) S-parameters of filters filled with ECCOSORB CR-110. Reflection (less than -20 dB) and transmission coefficients represented (linear dependency)

All filters fabricated according to the final technology route (Fig 4b) with a commercial filler ECCOSORB CR-110 showed a low level of reflection (about -20dB) over the entire measurement range, which is surpasses to modern analogues. The transmission parameter is smooth and has a linear dependence without significant fluctuations. All filters have very robust and repeatable characteristics. Further filters fabrication improvement can involve acquiring more accurate dependency of attenuation values on construction length for different fillers to investigate the possible reason of additional signal reflection present in the current filters.

## AUTHORS' CONTRIBUTIONS

A.A. Samoylov, A.I. Ivanov, and V.V. Echeistov contributed equally to this work.
Filters were made and characterized at the BMSTU Nanofabrication Facility (FMN Laboratory, FMNS REC, ID 74300).

## DATA AVAILABILITY

The data that support the findings of this study are available from the corresponding author upon reasonable request.